\documentclass[runningheads,a4paper]{llncs}

\usepackage{amssymb}
\usepackage{graphicx}

\usepackage{url}
\urldef{\mailsa}\path|{alfred.hofmann, ursula.barth, ingrid.haas, frank.holzwarth,|
\urldef{\mailsb}\path|anna.kramer, leonie.kunz, christine.reiss, nicole.sator,|
\urldef{\mailsc}\path|erika.siebert-cole, peter.strasser, lncs}@springer.com|
\newcommand{\keywords}[1]{\par\addvspace\baselineskip
\noindent\keywordname\enspace\ignorespaces#1}

\newcommand{\vp}{\varphi}
\RequirePackage{color}

\def \figr #1#2#3
{   \begin{figure}[ht]
        \centering\includegraphics[width=#2 \textwidth]{#1.eps}
        {\caption {#3}\label{#1}}
    \end{figure}
}   
\def \myfigures #1#2#3#4#5#6#7#8
{\begin{figure}[h!t]
    \centering\includegraphics[width=#2 \textwidth]{#1.eps}
        \hfill
    \centering\includegraphics[width=#6 \textwidth]{#5.eps}
        \parbox[t]{#4\textwidth}{\caption {#3}\label{#1}}
        \hfill
        \parbox[t]{#8\textwidth}{\caption {#7}\label{#5}}
\end{figure} }

\begin{document}
\mainmatter  

\title{Numerical study of magnetic flux in the LJJ model with double sine-Gordon equation}
\author{P.~Kh.~Atanasova,
\frame{T.~L.~Boyadjiev},
E.~V.~Zemlyanaya,
Yu.~M.~Shukrinov}
\authorrunning{Atanasova P.~Kh., Boyadjiev T.~L., Zemlyanaya E.~V., Shukrinov Yu.~M.}
\titlerunning{Stability analysis of magnetic flux in the LJJ model with 2SG-equation}
\institute{JINR, Dubna, Russia}
\maketitle

\begin{abstract}

The decrease of the barrier transparency in superconductor-insulator-superconductor (SIS)  Josephson junctions leads to
the deviations of the current-phase relation from the sinusoidal form. The sign of second harmonics is important
for many applications, in particular in junctions with a more complex structure like SNINS or SFIFS, where N is
a normal metal and F is a weak metallic ferromagnet. In our work we study the static magnetic flux
distributions in long Josephson junctions taking into account the higher harmonics in the Fourier-decomposition
of the Josephson current. Stability analysis is based on numerical solution of a spectral Sturm-Liouville
problem formulated for each distribution. In this approach the nullification of the minimal eigenvalue of this
problem indicates a bifurcation point in one of parameters. At each step of numerical continuation in parameters
of the model, the corresponding nonlinear boundary problem is solved on the basis of the continuous analog of
Newton's method. The solutions which do not exist in the traditional model have been found. The influence of
second harmonic on  stability of magnetic flux distributions for main solutions is investigated.

\end{abstract}

\section{Introduction}

Physical properties of magnetic flux in Josephson junctions (JJs) deserve the base of the modern superconducting
electronics. Tunnel SIS JJs  are known to be having the sinusoidal current
phase relation. However, the decrease of the barrier transparency in the SIS JJs leads
the deviations of the current-phase relation from the sinusoidal form
\cite{gki04}. We study the static magnetic flux distributions in the long JJs taking into account the second harmonic in the Fourier-decomposition
of the Josephson current. The sign of the second harmonic depends on physical applications under considering. It is important, in particular, in junctions like SNINS and SFIFS, where N is a normal metal and F is a weak metallic  ferromagnet \cite{ror01}. Interesting properties of long Josephson junctions with an arbitrarily
strong amplitude of second harmonic in current phase relation were
considered in \cite{goldobin07}.

Our purpose was to investigate an effect of the second harmonic accounting on the existence and stability magnetic flux distributions.
Below, the numerical scheme and results of our stability analysis  are demonstrated.

\section{Mathematical statement of the problem}

For a sufficiently wide class of JJ the superconducting Josephson current as a function of magnetic flux $\vp$ (phase difference of superconductors wave functions) can be represented as a sine series \cite{l85}:
\begin{equation}\label{i_s_full}
    I_S = I_c \sin \varphi + \sum_{m=2}^{\infty} I_m \sin m
    \varphi\,.
\end{equation}
Using only first two terms of this expansion one can show \cite{htkt00} that the distribution of the magnitude $\vp(x)$ along $x$-axis of the junction in the static regime \cite{l85} satisfies the double {sine}-Gordon equation (2SG).

\begin{equation}\label{2sg}
    -\varphi\,'' + a_1 \sin \varphi + a_2 \sin 2\varphi -\gamma = 0\,, x \in (-l;l)\,.
\end{equation}
 Here and below the prime means a derivative with respect to the coordinate $x$. The magnitude $\gamma$ is the external current, $l$ is the semilength of the junction,  $a_1$ and $a_2$ are parameters corresponding to $I_c$ and $I_2$ in (\ref{i_s_full}) respectively. They depend on the preparation technology of junctions \cite{gki04,bk03}.  All the magnitudes are dimensionless.

In the case of in-line geometry of the junction the boundary conditions for (\ref{2sg}) have the form
\begin{equation}\label{bound2sg}
    \varphi\,'(\pm l) = h_e,
\end{equation}
where $h_e$ is  external magnetic field.

From the mathematical viewpoint the transfer of the junction into dynamical regime \cite{l85} means \cite{galfil_84,pbvzpc07} a  stability loss  (bifurcation) of all static solutions $\vp(x)$ of (\ref{2sg}), (\ref{bound2sg}) at the parameters $\gamma$ or $h_e$ variation. Our stability analysis
of $\vp(x,p)$ was based on numerical solution of the corresponding Sturm-Liouville problem
\begin{equation}
\label{slp}
   -\psi\,'' + q(x)\psi = \lambda \psi, \quad
    \psi\,'(\pm l) = 0
\end{equation}
with a potential $q(x) = a_1 \cos \varphi + 2 a_2 \cos 2 \varphi$.

The minimal eigenvalue  $\lambda_0(p) > 0$ corresponds the stable solution. In case $\lambda_0(p) < 0$ solution $\vp(x,p)$ is unstable. The case $\lambda_0(p) = 0$ indicates the bifurcation with respect to one of parameters $p = (l,a_1,a_2,h_e,\gamma)$.

\section{Numerical method}

Numerical solving of the boundary problem (\ref{2sg}),(\ref{bound2sg}) was performed on the basis of the Continuous analog of Newton's method \cite{pbvzpc07}. At each  Newtonian iteration the corresponding linearized problem was solved using three-point
Numerov's finite-difference approximation of the fourth  order accuracy \cite{numerov}. The discretization of  the Sturm-Liouville  problem (\ref{slp}) was realized with the help of standard second order finite-difference formulae. The calculation of the first several eigenvalues of the corresponding algebraic $3$-diagonal
problem was performed applying the standard subroutine from the package  EISPACK.
Details of numerical scheme are described in \cite{azbsh10}

\figr{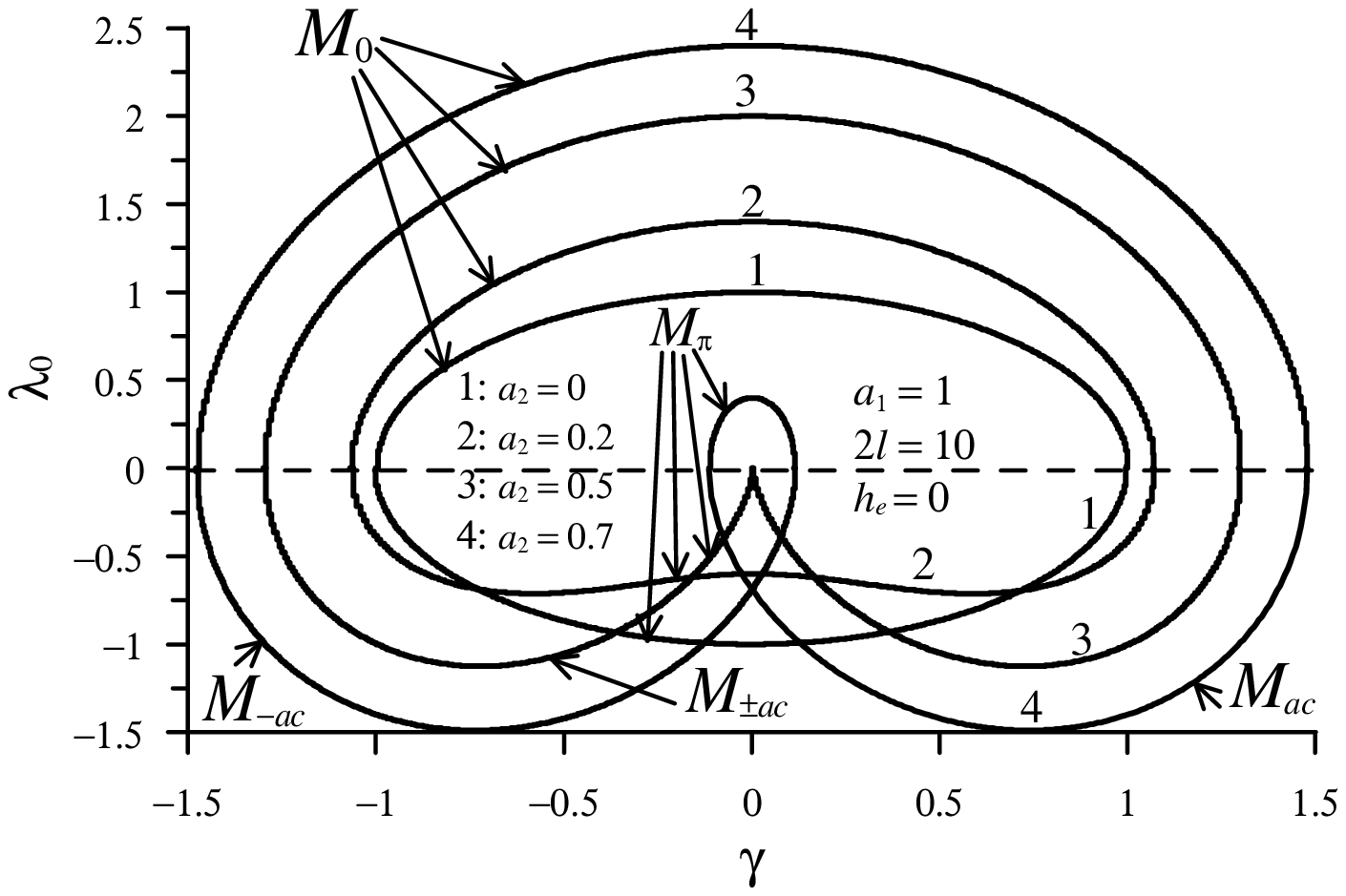}{0.7}{Change of $\lambda_0(\gamma)$ for CS with increase of the coefficient $a_2$ in the interval $a_2 \in [0;0.7]$ at $h_e = 0$, $a_1 = 1$, $2l = 10$.}

\figr{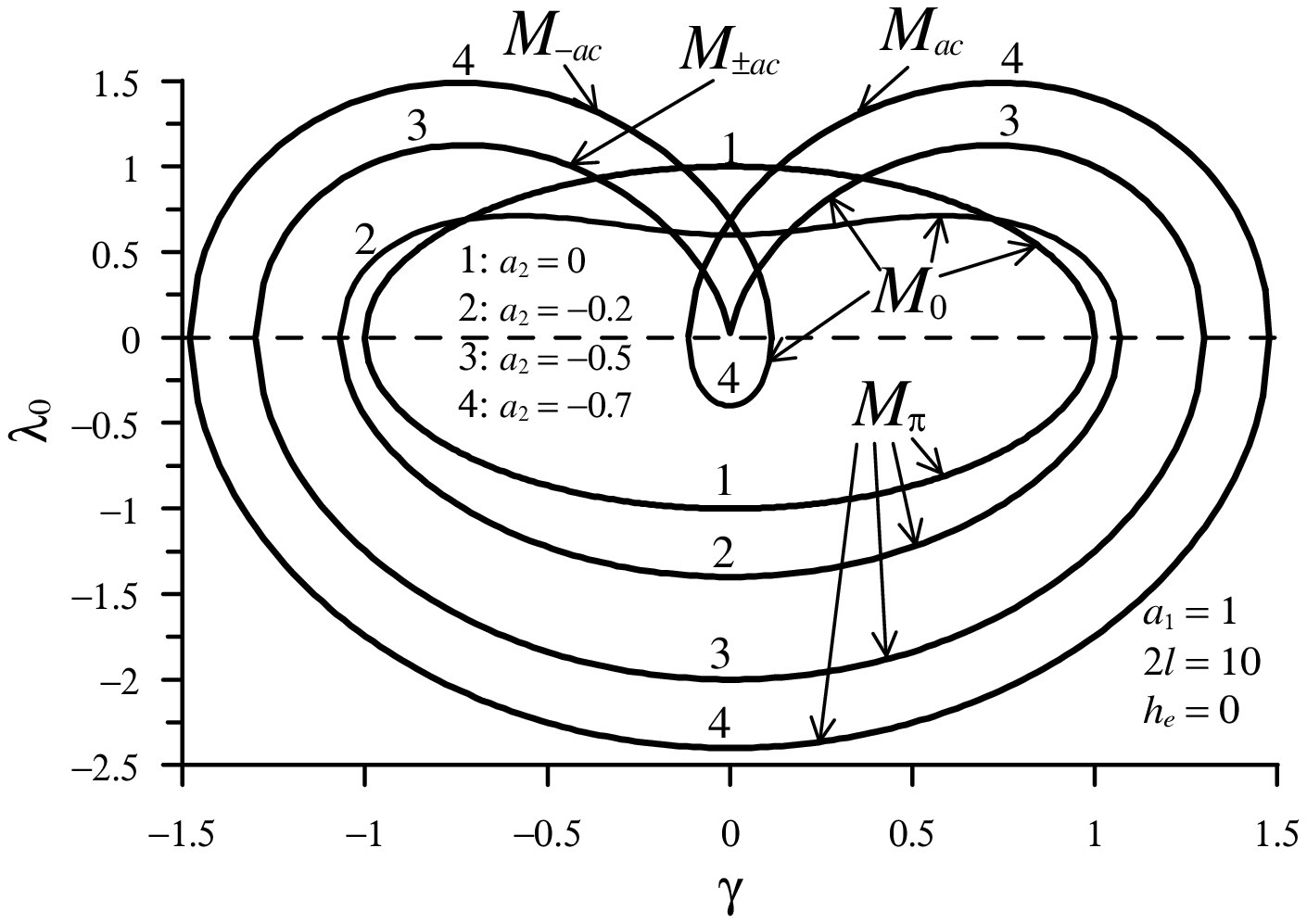}{0.7}{The same as on Fig. 1 but for $a_2 \in [-0.7;0]$.}

\section{Numerical results and conclusions} \label{numerical_results}

Let us start with the {\it trivial solutions} of (\ref{2sg}).
In the ``traditional'' case $a_2 = 0$ two trivial solutions $\varphi = 0$  and $\varphi = \pi$ (below they are denoted by $M_{0}$ and $M_{\pi}$ respectively)  are known  at $\gamma = 0$ and $h_e = 0$. Accounting of the second harmonic $a_2 \sin 2\varphi$ leads appearing two additional solutions  $\varphi = \pm \arccos (-a_1/2a_2)$ (denoted as $M_{\pm ac}$). The corresponding $\lambda_0$ as functions of  2SG-equation coefficients  have the form   $\lambda_0[M_0] = a_1 + 2 a_2$, $\lambda_0[M_{\pi}] = -a_1 + 2 a_2$ and $\lambda_0[M_{\pm ac}] = (a_1^2 - 4 a_2^2)/2 a_2$. The exponential stability of these constant solutions (CS) is determined by the signs of the parameters $a_1$ and $a_2$ and by its ratio $a_1/a_2$ \cite{azbsh10}.



\myfigures{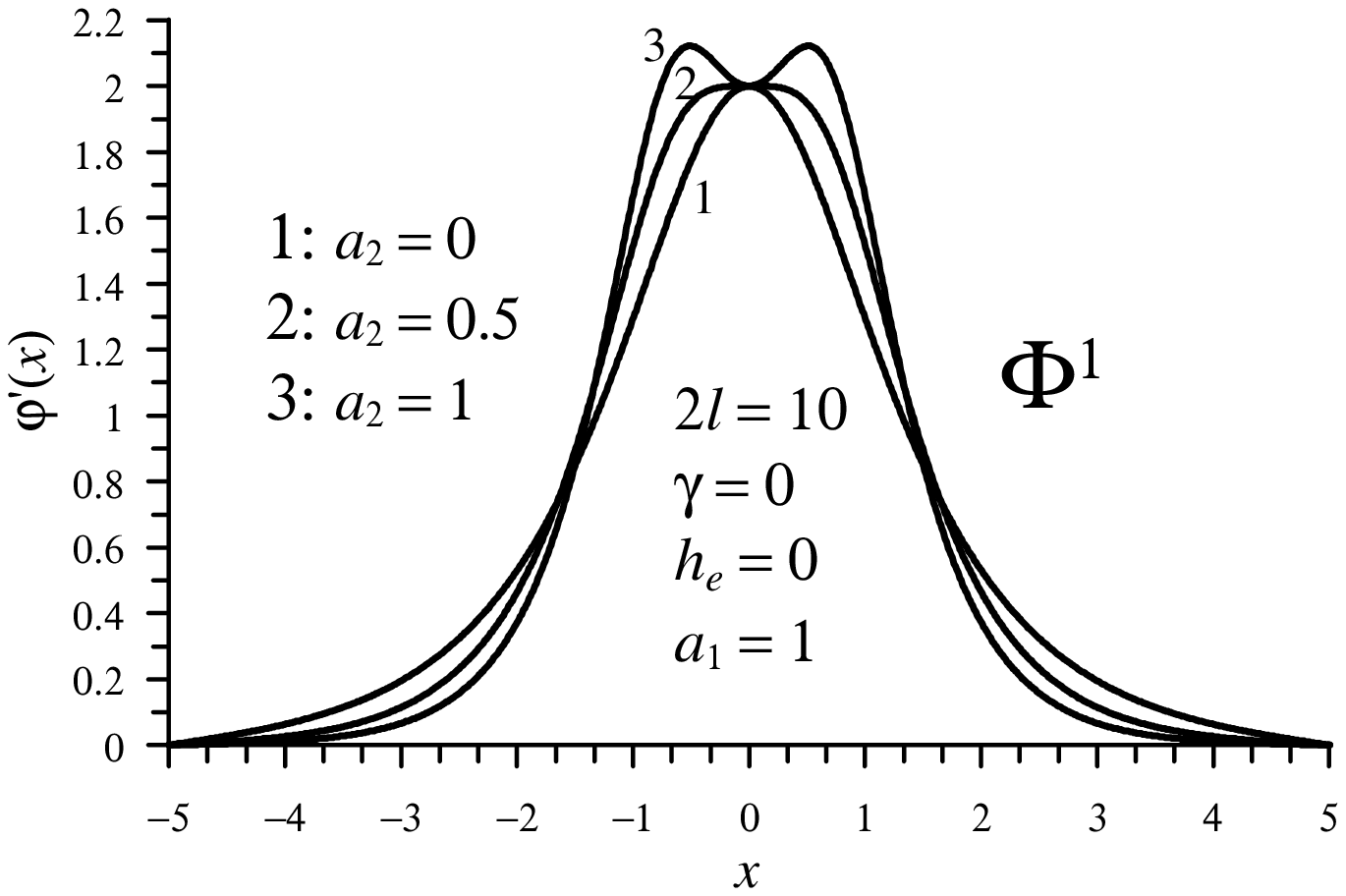}{0.48}{Internal magnetic field of the fluxon $\Phi^1$ at $\gamma = 0$, $h_e = 0$ and $2l = 10$ when the parameter $a_2 \geq 0$ increases.}{0.49}
{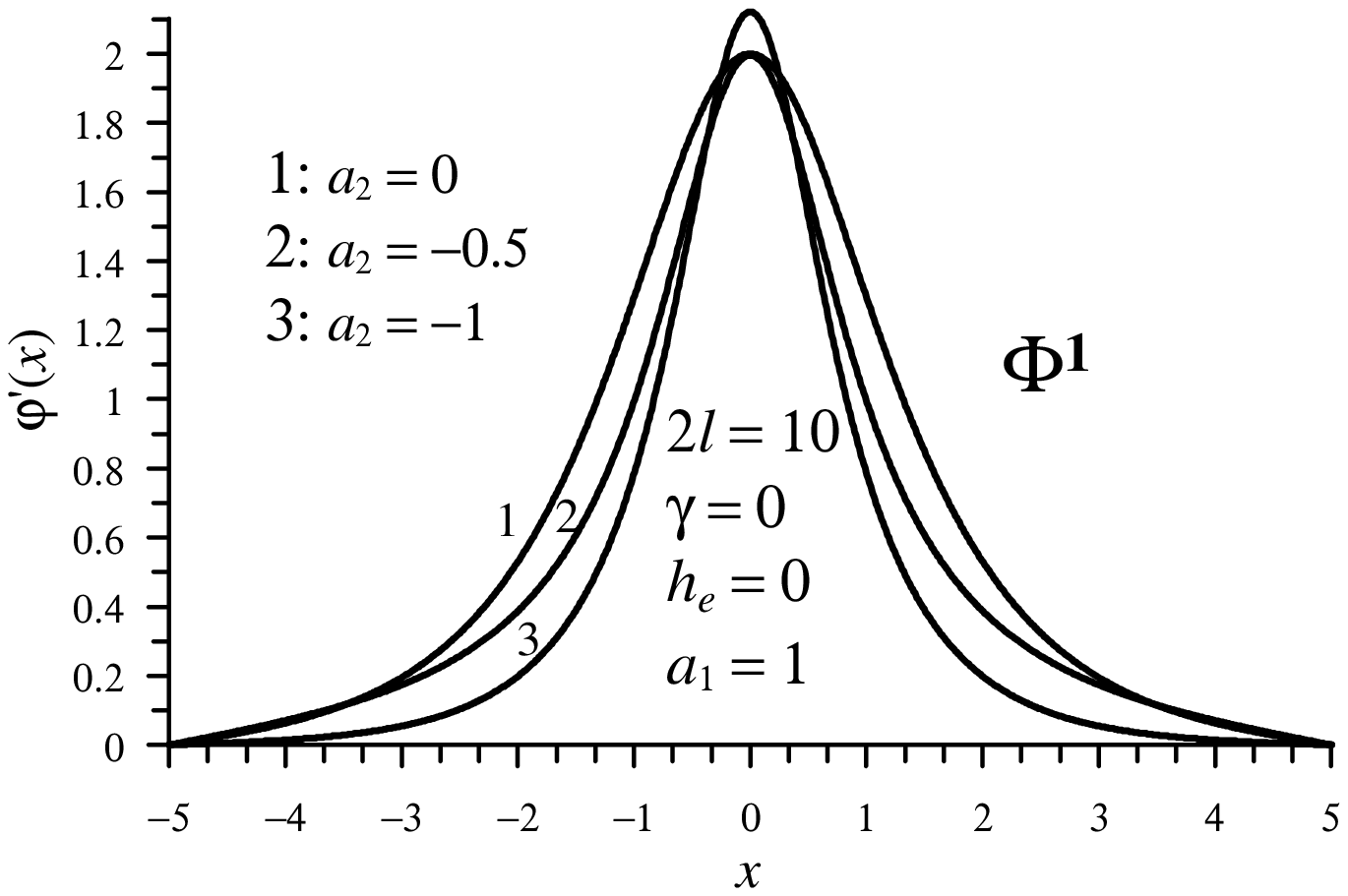}{0.48}{
The same as on Fig. 3 but for decreasing $a_2 \leq 0$.
}{0.46}


\myfigures{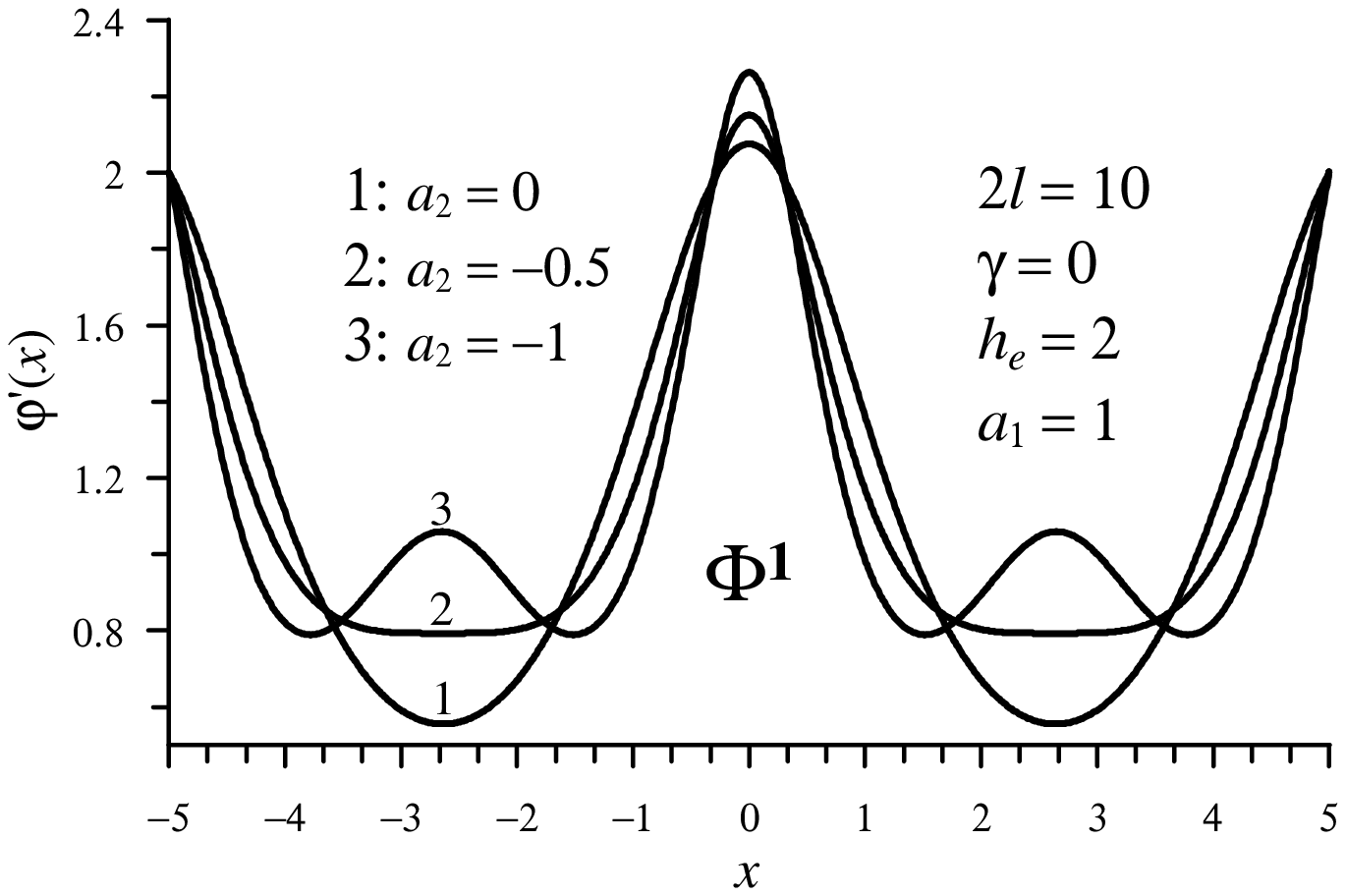}{0.48}{Internal magnetic field of the fluxon $\Phi^1$ at $\gamma = 0$, $h_e = 2$ and $2l = 10$ when the parameter $a_2 \leq 0$ decreases.}{0.49}
{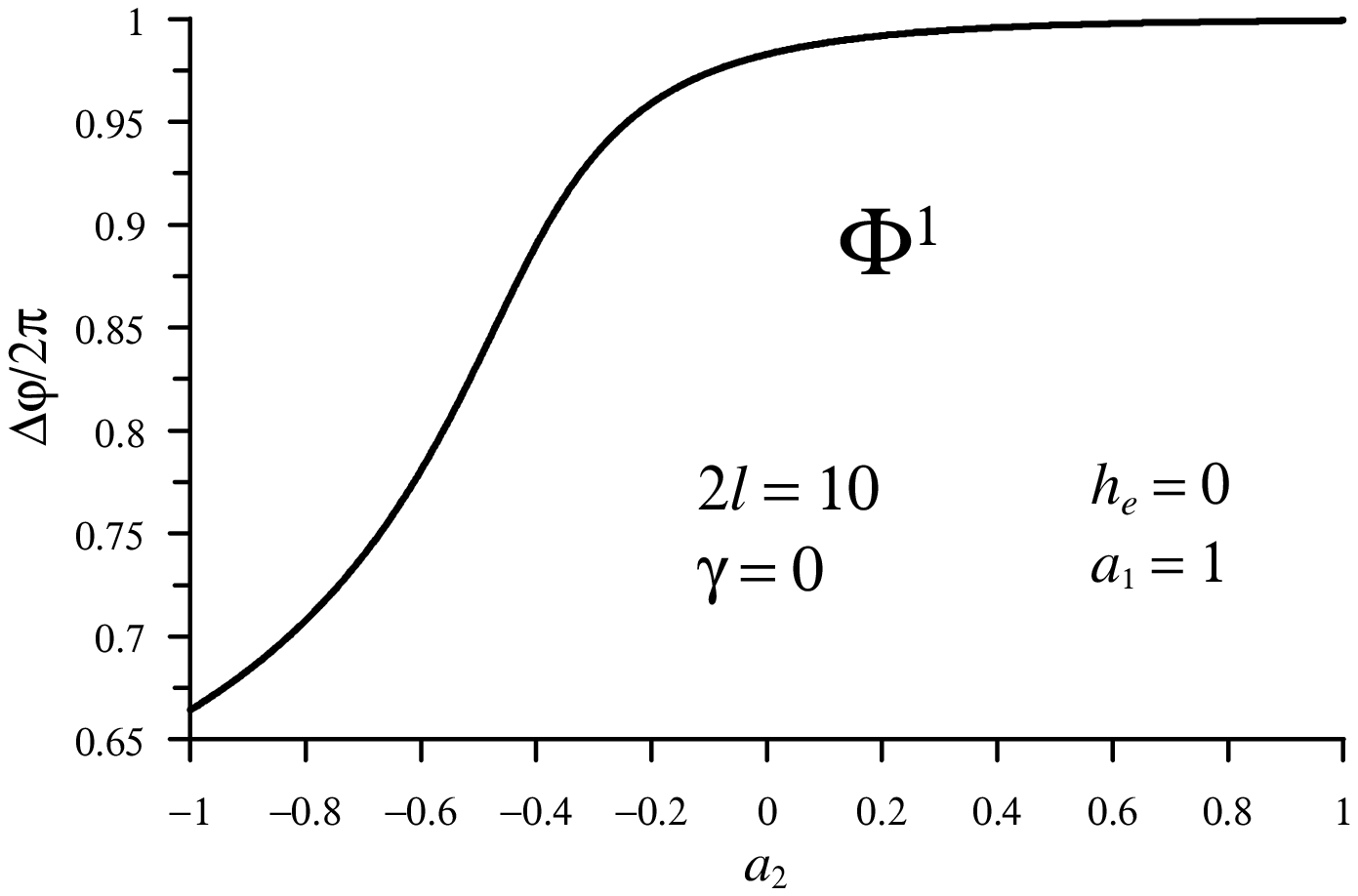}{0.48}{Full magnetic flux in dependence on the parameter $a_2 \in [-1;1]$  at $h_e = 0$, $\gamma = 0$, $2l = 10$ for $\Phi^1$.}{0.46}

The dependencies of $\lambda_{0}$ on the external current $\gamma$ for CS at several positive values of $a_2$
 are demonstrated in Fig.\ \ref{ev0_gamma_he0_cs_a1_1_a2_0_02_05_07_paper}. Arising of the stable states $M_{\pi}$ by the external current $\gamma$ at $a_2 > 0.5$ is shown.

When $a_2 < -0.5$ the stable solution $M_{0}$  disappears and other stable constant solutions $M_{\pm ac}$ arise. This transition is seen  in Fig.~\ref{ev0_gamma_he0_cs_a1_1_a2_0_-02_-05_-07_paper}.

Excepting CS, the 2SG equation is known to be supporting {\it fluxon solutions}. The fluxons play a significant role in the JJ physics. Different distributions of magnetic flux in JJ are considered in the review \cite{pbvzpc07}. At small external fields $h_e$ such distributions are fluxon $\Phi^1$, antifluxon $\Phi^{-1}$ and their bound states $\Phi^1\Phi^{-1}$ and $\Phi^{-1}\Phi^1$ . As external magnetic field $h_e$ is growing, more complicated stable fluxon and bound states appear: $\Phi^{\pm n}$ and $\Phi^{\pm n}\Phi^{\mp n}$ $(n=1,2,3,...)$.

Let us  compare  some basic physical characteristics of one-fluxon solution  $\Phi^1$ in our model (\ref{2sg}),(\ref{bound2sg}) with the traditional model ($a_1 = 1$, $a_2 = 0$). In Fig.\ \ref{sols_f1_a1_1_a2_0_05_1_paper}  the deformation of the $\vp\,'(x)$ under influence of the parameter $a_2 \in [0;1]$ is demonstrated. At $a_2 = 0.5$ the curve of internal magnetic field $\vp\,'(x)$ has a  plateau in a neighborhood  of the center $x = 0$. Further increase of the parameter $a_2$ leads to a formation of two maximums of the magnetic field. Thus, taking account of the coefficient $a_2$ leads to qualitative change of the form of fluxon distribution $\Phi^1$. Such a change is not seen with a decrease in parameter $a_2$ when $h_e = 0$ (Fig.~\ref{sols_f1_a1_1_a2-_paper}). In the case of sufficiently large \textit{\textcolor[rgb]{0.75,0.75,0.75}{a}} external magnetic field $h_e$ one can observe a similar qualitative  deformation in the local minimums regions for  $a_2 < 0$ (see Fig.\ \ref{sols_f1_a1_1_a2_0_-05_-1_he_2_paper}).

With change of the coefficient $a_2$ the number of fluxons \cite{pbvzpc07}
$$ N(p) = \frac{1}{2l \pi} \int\limits_{-l}^l \varphi(x)\,dx,$$
corresponding to the distribution $\Phi^1$ is conserved i.e. $ {\partial N}/{\partial a_2 } = 0$. Here we have a value $N[\Phi^1] = 1$.

When $a_2$ is growing the full magnetic flux \cite{pbvzpc07} $\Delta\varphi(p) = \varphi(l) - \varphi(-l)$
for this solutions tends to $2\pi$, see Fig.~\ref{flux_a2_a1_1_2l10_he0_g0_F1}.

The value of the magnetic flux $\vp(x)$ in the middle of the interval does not change and $\vp(0) = \pi$.

\vspace*{.2cm}

In this paper we only focused on stability analysis of constant solutions and one-fluxon solutions in dependence on the $a_2$ contribution. Investigation of another classes of solutions of 2GS-equation is the point of our further research.

\subsubsection*{Acknowledgments.}

%
The authors are thankfull to I.V.Puzynin and T.P. Puzynina for usefull discussions, valuable remarks and for the support of this work.
The work of  P.Kh.A. is partially financed by the Program for collaboration of JINR-Dubna and  Bulgarian scientific center ``JINR -- Bulgaria''.
E.V.Z.  and Yu.M.Sh. are grateful  to  RFFI  (grants 09-01-00770-a and 08-02-00520-a, correspondingly) for a partial financial support.


\end{document}